\pdfoutput=1

\documentclass[11pt]{article}

\usepackage{ACL2023}

\usepackage{times}
\usepackage{latexsym}
\usepackage{amsmath}
\usepackage{amssymb}
\usepackage{listings}
\usepackage{booktabs}

\usepackage[T1]{fontenc}

\usepackage[utf8]{inputenc}

\usepackage{microtype}

\usepackage{inconsolata}

\usepackage{graphicx}
\usepackage{multirow}
\usepackage[normalem]{ulem}
\useunder{\uline}{\ul}{}

\title{Rewriting the Code: A Simple Method for Large Language Model Augmented Code Search}

\author{{\bf Haochen Li}\ \ \ \ 
    {\bf Xin Zhou\thanks{\ \ Corresponding author}}\ \ \ \ 
    {\bf Zhiqi Shen}  \\
     Nanyang Technological University, Singapore \\
     \texttt{\{haochen003, xin.zhou, zqshen\}@ntu.edu.sg
     }}

\begin{document}
\maketitle
\begin{abstract}
In code search, the Generation-Augmented Retrieval (GAR) framework, which generates exemplar code snippets to augment queries, has emerged as a promising strategy to address the principal challenge of modality misalignment between code snippets and natural language queries, particularly with the demonstrated code generation capabilities of Large Language Models (LLMs).
Nevertheless, our preliminary investigations indicate that the improvements conferred by such an LLM-augmented framework are somewhat constrained. This limitation could potentially be ascribed to the fact that the generated codes, albeit functionally accurate, frequently display a pronounced stylistic deviation from the ground truth code in the codebase. 
In this paper, we extend the foundational GAR framework and propose a simple yet effective method that additionally \underline{Re}writes the \underline{Co}de (ReCo) within the codebase for style normalization. 
Experimental results demonstrate that ReCo 
significantly boosts retrieval accuracy across sparse (up to 35.7\%), zero-shot dense (up to 27.6\%), and fine-tuned dense (up to 23.6\%) retrieval settings in diverse search scenarios. 
To further elucidate the advantages of ReCo and stimulate research in code style normalization, we introduce Code Style Similarity, the first metric tailored to quantify stylistic similarities in code. 
Notably, our empirical findings reveal the inadequacy of existing metrics in capturing stylistic nuances. The source code and data are available at \url{https://github.com/Alex-HaochenLi/ReCo}.
\end{abstract}

\section{Introduction}

\begin{figure}[ht]
\centering
\includegraphics[width=0.98\columnwidth, trim=0 270 0 0, clip]{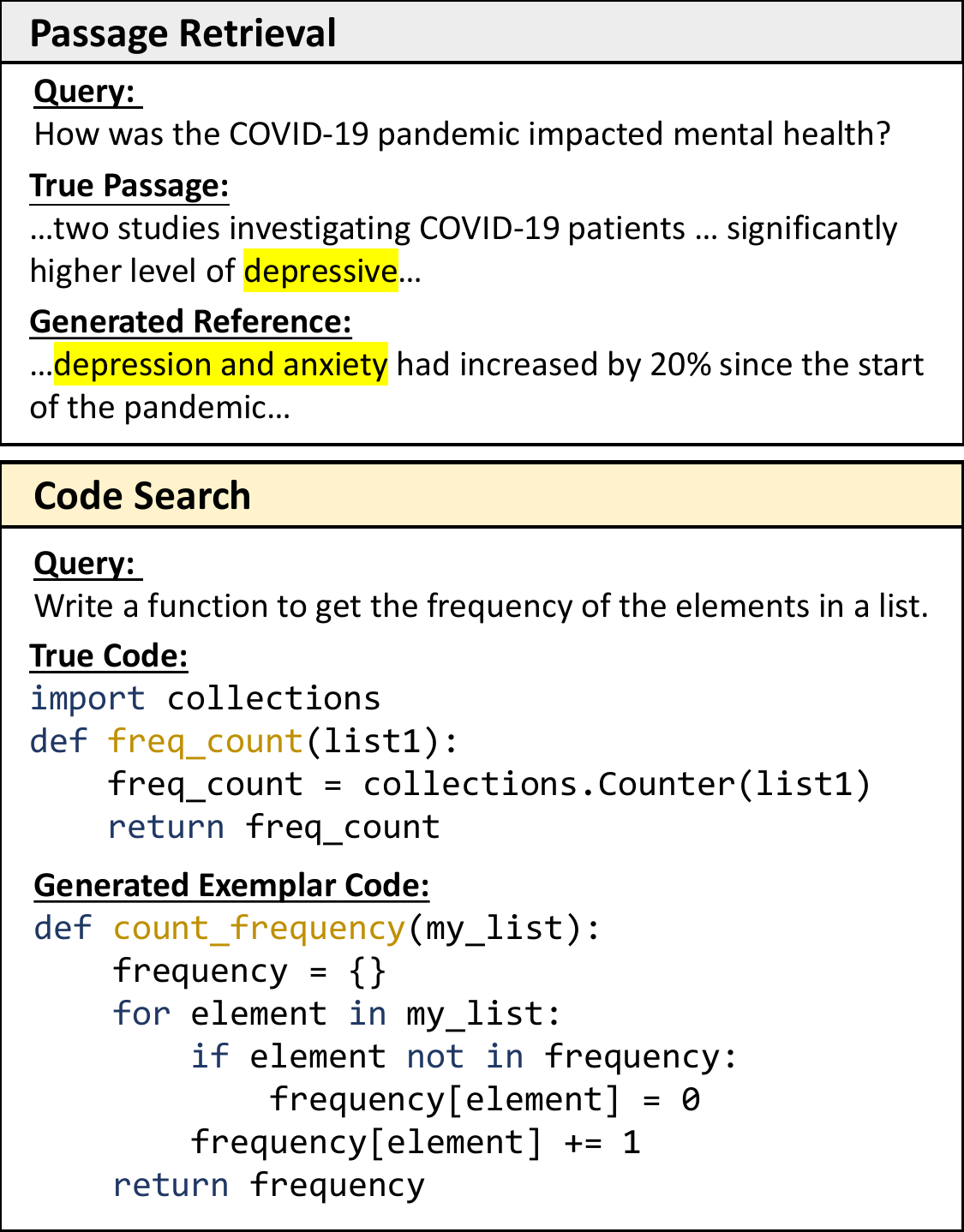}
\caption{Comparison of GAR between passage retrieval and code search. In passage retrieval, the truth (yellow) is included in the generated content. In code search, despite the generated exemplar code satisfies the description of the query, it exhibits noticeable dissimilarity to the true code.}
\label{fig1}
\end{figure}

\begin{figure*}[t]
\centering
\includegraphics[width=1.98\columnwidth, trim=0 545 420 0, clip]{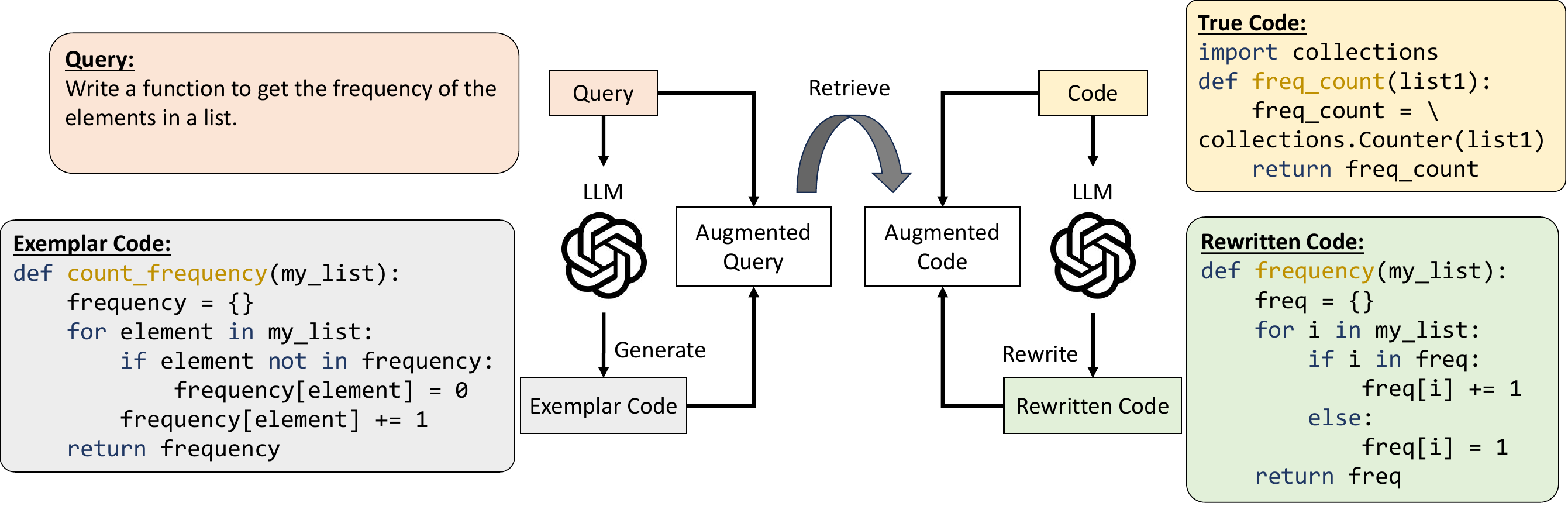}
\caption{An illustration of the ReCo method. 
It initially prompts LLMs to generate exemplar codes based on the search query. Subsequently, the original query and these exemplar codes are synthesized to formulate an augmented query. Analogously, the rewritten codes, produced by the LLMs, are merged with the original code, thereby creating a candidate for retrieval. The example delineated in this figure aligns with the one depicted in Fig.~\ref{fig1}.}
\label{fig2}
\end{figure*}

Code search, aimed at retrieving the most semantically relevant code snippets from a codebase according to a specified natural language query, is a common activity that plays an important role in software development \cite{nie2016query, shuai2020improving}. 
Retrieving and reusing analogous code fragments from large-scale codebases like GitHub can enhance productivity significantly. 

Despite both being sequences of words, matching code queries and natural language queries is challenging as they share few grammatical rules, causing them to fall into two distinct modalities. 
This grammatical distinction results in limited word overlap, significantly hampering the application of sparse retrieval systems in code search.
On the other hand, in dense retrieval systems, the alignment of query and code representations during the training phase assists in alleviating the challenge \cite{softinfonce}.  As a result, these systems are capable of encapsulating potential semantic correlations between terminologies employed in programming languages and those in natural languages. However, this potential association becomes challenging to capture if two terminologies rarely manifest together within a query-code pair. 

To bridge this gap, one possible solution is to transform the data from one modality to the other. This could involve either generating exemplar codes based on the query or summarizing the functionality of codes in the codebase. Given that natural language queries in code search are often short and ambiguous \cite{mao2023self,rahman2021systematic},  
our research concentrates on the former solution, referred as Generation-Augmented Retrieval (GAR) \cite{gar}. 
GAR has demonstrated competitive performance in question answering and passage retrieval. In these NLP tasks, a language model is adopted to generate references based on the query to augment it.
Similarly, we could use a language model to generate exemplar code snippets that realize the functionalities described in the query. Then the query and exemplar codes are combined to be fed into the retrieval system.
With many LLMs demonstrating great intelligence in precisely writing codes \cite{llama,llama2,gpt4, nl2codesurvey}, performing GAR with LLMs becomes a promising approach for code search.

However, from our preliminary studies, the improvement in performance brought by GAR using LLMs is limited, especially with the high computational cost of LLMs. 
We argue that answer format influences the performance of GAR on question answering and code search.  
In question answering, the correct answer to the question is often unique and can be expressed in limited forms. The generated contents from LLMs, if correct, are usually in the exact same form as the answer. 
As highlighted in Fig.~\ref{fig1}, the matching word ``depressive'' appears in the reference. 
On the other hand, code snippets with the same functionality can have diverse formulations, which lowers the chance of matching the code in the codebase, and thus leads to minor improvement of GAR in code search.
As shown in Fig.~\ref{fig1}, the true code uses Python built-in function \texttt{Counter} to count the number of elements in a list, while the exemplar code snippet does it manually.

To address the mismatch of the generated and ground truth code snippets, we build upon GAR and propose a simple yet effective framework that additionally \underline{Re}writes and the \underline{Co}de (ReCo) in the codebase. 
As shown in Fig.~\ref{fig2}, after rewriting, the style of codes in the codebase are normalized by LLMs to align with the exemplar code, thereby facilitating the retrieval. 
We evaluate ReCo on several code search models across various search scenarios, including coding challenge competence, online programming community, and general programming problems in Python and Java. Experimental results show that ReCo could significantly boost the performance of sparse retrieval systems (up to 35.7\%) and dense retrieval systems in both zero-shot (up to 27.6\%) and fine-tuning (up to 23.6\%) settings. 

Furthermore, we propose a novel evaluation metric, dubbed Code Style Similarity, to quantitatively measure the disparity in code style. Our metric validates ReCo's capability in aligning the style of code within the codebase with that of code generated by LLMs. 
Conventional metrics like BLEU \cite{bleu} and CodeBLEU \cite{codebleu} are deemed less appropriate as they calculate similarity based on exact-matched tokens of the given two code snippets. 
In contrast, Code Style Similarity evaluates style from three distinct perspectives: variable naming, API invocation, and code structure, based on edit distance \cite{editdistance}. 
Our experiments show that Code Style Similarity exhibits superior explanatory power than existing metrics in measuring the style deviation of code from the dataset and that generated from LLM.

\section{Related Works}
\paragraph{Code Search Models}
The development of code search models could be split into three stages. Traditional methods, also denoted as sparse retrieval, employ information retrieval techniques to match words between queries and codes \cite{hill2011improving,yang2017iecs,satter2016search}. As we mentioned before, since programming languages and natural languages share few grammatical rules, these methods often suffer from vocabulary mismatch problems \cite{mcmillan2011portfolio}. 
Then, neural models became popular \cite{cqil,cambronero2019deep,gu2018deep,codesearchnet}.  
They all employ a framework where queries and codes are encoded by neural encoders separately into a joint representation space. 

Recently, transformer-based pre-trained models significantly outperformed previous methods, since they can be trained on large-scale unlabelled corpus with self-supervised pre-training tasks. Many novel pre-training tasks are proposed to let models have a better general understanding of codes \cite{guo2020graphcodebert, li2022coderetriever, scoder, cocosoda}. 
For instance, CodeBERT \cite{codebert} utilizes masked language modeling and replaced token detection. 
CodeT5 \cite{codet5} focuses on generative tasks through bimodal dual generation. UniXcoder \cite{unixcoder} integrates the aforementioned generative and understanding pre-training tasks. CodeT5+ \cite{codet5+} employs the same architecture as CodeT5 and pre-trains it with span denoising, causal language modeling, contrastive learning, and text-code matching from both unimodal code data and bimodal code-text data.

\paragraph{Large Language Models}
As the model parameters and size of training corpora of those transformer-based pre-trained models scale up to billions, they appear to demonstrate remarkable intelligence in understanding and generating codes. As a milestone, Codex \cite{codex} with 12 billion parameters indicates the beginning of the Code LLM era. Meanwhile, there are a number of powerful Code LLMs proposed \cite{nl2codesurvey}, though most of them are not publicly accessible. Recently, ignited by OpenAI's ChatGPT \cite{chatgpt}, a bunch of excellent open-sourced models also contribute to the thriving of Code LLMs \cite{codellama,codegen,wizardcoder}. Among them, Code LLaMA \cite{codellama} has attracted significant attention because it is a collection of efficient Code LLMs ranging from 7B to 34B parameters. At the same time, some LLMs that are not specifically trained for code exhibit surprising abilities in code intelligence as well, such as GPT3.5 \cite{chatgpt} and LLaMA \cite{llama,llama2}. 
This can be attributed to the inclusion of code snippets in the unlabeled training corpus.

\paragraph{LLMs for Retrieval}
While LLMs are designed for token generation, their direct application to retrieval tasks such as code search is not suitable.
Indeed, there have been attempts to amalgamate the search query and all the candidates together as input, subsequently requesting the LLMs to rank the candidates within the input \cite{qin2023large}. However, the constraint on input sequence length impedes its applicability to large-scale retrieval tasks.

One indirect way is to ask LLMs to generate some references and expand the search query with them. This framework, denoted as Generation-Augmented Retrieval, has been proven effective in both question answering and passage retrieval \cite{gar,hyde,query2doc}. \citet{gar} is the first work to propose GAR in question answering. HyDE \cite{hyde} evaluates GAR in passage retrieval under zero-shot setting. query2doc \cite{query2doc} extends GAR to fine-tuning. 
Our research findings suggest that the Generation-Augmented Retrieval (GAR) method does not substantially enhance the efficiency of code search, primarily due to the significant stylistic difference between exemplar code and true code.

\paragraph{Code Generation Evaluation Metrics}
A suitable automatic evaluation metric is vital to the growth of code generation. It is used to measure the lexical similarity between the generated hypothetical code and the true reference code. 
Initially, metrics such as BLEU \cite{bleu} and ROUGE \cite{rouge}, originally designed for machine translation, were utilized in the realm of code generation. However, subsequent scholarly discourse posits that these metrics overlook the syntactic and semantic nuances inherent to code.
Hence, to consider those features, CodeBLEU \cite{codebleu} adds terms that calculate Abstract Syntax Tree similarity and data-flow similarity. CrystalBLEU \cite{crystalbleu} sets weights for tokens according to their frequency. They find that high-frequency tokens are often meaningless hence assigning lower weights. These metrics are widely adopted in code generation evaluation, yet they are not suitable for measuring the style difference between two codes due to shared syntactic verbosity.

\section{Methodology}
\subsection{Preliminaries}
Code search aims to retrieve code snippets that are semantically most pertinent to a specified query. 
Given a search query $q$ and a code snippet $c$ in the fixed codebase, an encoder $G$ is used to map the query and the code to a shared representation space. We calculate the similarity between query and code by dot product, which could be formulated as:

\begin{equation}
\label{eq:sim}
    sim(q,c) = \langle G(q), G(c) \rangle = \langle \mathbf{v}_q, \mathbf{v}_c \rangle,
\end{equation}
where $\mathbf{v}_q$ and $\mathbf{v}_c$ are representation vectors of $q$ and $c$, respectively. Finally, codes in the codebase are ranked according to the similarity score. Note that in a code search system, code representations $\mathbf{v}_c$ can be calculated and stored in advance. 

\subsection{ReCo}
Building on GAR, ReCo not only generates exemplar codes based on the query but also rewrites the codes in the codebase. 

\paragraph{Generating and Rewriting Code}
First, we elucidate the process of generating exemplar codes. Given a query $q$, we employ few-shot prompting (a.k.a in-context learning) \cite{gpt3} to generate an exemplar code snippet. The prompt consists of an instruction \textit{``Please generate a Java/Python code snippet according to the given description.''} and $K$ randomly sampled query-code pairs from the training set. In this paper, we set $K=4$. 
The instruction and in-context samples are denoted as \textsc{gen}, enabling us to derive the exemplary code $c_q$ as follows:

\begin{equation}
    c_q = \mathrm{LLM}(q, \mathrm{\textsc{gen}}).
\end{equation}

In the procedure of rewriting the code $c$, we initially summarize the code into a natural language description, represented as $q_{sum}$. This can be achieved by changing the instruction in \textsc{gen} to \textit{``What is the main purpose of the Java/Python code snippet?''}. We denote it as \textsc{sum}. Subsequently, similar to generating exemplar codes, we consider $q_{sum}$ as the query $q$\footnote{The process of rewriting is implemented via a summarize-then-generate approach, as we have observed that merely instructing LLMs to rewrite the original codes does not result in significant alterations.}. The entire process leading to the acquisition of the rewritten code, denoted as $c_c$, is as follows:

\begin{equation}
    q_{sum} = \mathrm{LLM}(c, \mathrm{\textsc{sum}}),
\end{equation}
\begin{equation}
    c_c = \mathrm{LLM}(q_{sum}, \mathrm{\textsc{gen}}).
\end{equation}

Detailed examples are provided in Appendix~\ref{app:prompt} to further elucidate the prompt.

\paragraph{Sparse Retrieval}
Query $q$ and code $c$ are appended with exemplar code $c_q$ and rewritten code $c_c$ in a sparse retrieval system, respectively. Since we could generate multiple code snippets as augmentation, to retain the original semantics of $q$ and $c$, we simply repeat them for $N$ times which is equal to the number of augmented codes. Take the query as an example, the augmented search query $q^+$ could be expressed as:

\begin{equation}
    q^+ = \mathrm{concat}(\{q\} \times N, \{c_{q1},c_{q2},\ldots, c_{qN}\}).
\end{equation}

Similarly, we could get the augmented code $c^+$. In application, $q^+$ is fed to the sparse retrieval system as the search query and $c^+$ are candidates in the codebase.

\paragraph{Dense Retrieval}
InfoNCE loss \cite{infonce} is widely adopted in fine-tuning because it can pull together the representations between the query and its corresponding code while pushing away the representation of negative codes \cite{softinfonce,li2022exploring}. During training, we take other in-batch codes as negative samples for a query \cite{cosqa}. With augmented query $q^+$ and augmented code $c^+$, InfoNCE loss $\mathcal{L}$ can be described as:

\begin{equation}
\small
    \mathcal{L} = -\mathbb{E} \left [\log \frac{\exp (\mathbf{v}_{qi}^+\cdot \mathbf{v}_{ci}^+)}{\exp (\mathbf{v}_{qi}^+\cdot \mathbf{v}_{ci}^+)+\sum_{j\neq i}^n\exp (\mathbf{v}_{qi}^+ \cdot \mathbf{v}_{cj}^+)} \right],
\end{equation}
where $n$ is the batch size, $\mathbf{v}_{q}^+$ and $ \mathbf{v}_{c}^+$ are augmented representations of $q^+$ and $c^+$, respectively. 

For augmented representations, we calculate the expectation of all the generated content according to the chain rule. Take the exemplar code as an example, we have:

\begin{equation}
    \mathbb{E} [\mathbf{v}_{cq}] = \mathbb{E}[G(c_q)] = \mathbb{E}[G(\mathrm{LLM}(q, \textsc{gen}))].
\end{equation}

Here we assume the distribution of $\mathbf{v}_{cq}$ is uni-modal since the preferred style of LLM is consistent when generating codes. Then, we employ average pooling between the representation of $\mathbf{v}_{cq}$ and $\mathbf{v}_{q}$ to get the augmented representation $\mathbf{v}_{q}^+$. The total process can be described as:

\begin{equation}
    \mathbf{v}_{q}^+ = \frac{1}{2N} \left ( N \cdot G(q) + \sum_{c_q \sim \mathrm{LLM}(q, \textsc{gen})} G(c_q) \right ),
\end{equation}

where $N$ is the number of exemplar codes. Similarly, we can get the augmented representation $\mathbf{v}_{c}^+$ of each code. During evaluation, $\mathbf{v}_q$ and $\mathbf{v}_c$ in Eq.\eqref{eq:sim} are replaced by $\mathbf{v}_q^+$ and $\mathbf{v}_c^+$.

\paragraph{Theoretical Insights}
We offer theoretical insights to differentiate GAR and our proposed ReCo. Each code in the codebase is an implementation of a specific query, which we denote as $c\sim P(q)$. Here $P$ denotes a real-world distribution between queries and codes. LLM also defines a probability distribution over queries. Thus, exemplar codes can be considered to follow $c_q\sim \mathrm{LLM}(q)$. We could find that $c$ and $c_q$ are sampled from two different distributions given $q$. 
This accounts for the occasional divergence between the true code and the exemplar code for the same query, as illustrated in Fig.~\ref{fig1}.

The rewritten code follows $c_c\sim \mathrm{LLM}(q_{sum})$. If the query $q$ and code $c$ are identical in semantics and $q_{sum}$ correctly reflect the functionality of code $c$, we could approximate the distribution of $\mathrm{LLM}(q_{sum})$ as $\mathrm{LLM}(q)$. Once the exemplar code and the rewritten code are both sampled from $\mathrm{LLM}(q)$, the expectation of LLM-generated content becomes more similar, which is reflected in the style of generated codes.

\section{Code Style Similarity}
To quantitatively measure the style difference among codes, we propose a novel evaluation metric, dubbed Code Style Similarity (CSSim). To the best of our knowledge, this is the first metric addressing the similarity between two codes from a  stylistic perspective. Indeed, there are several evaluation metrics widely adopted in code generation or translation to measure semantic similarity like BLEU and CodeBLEU. Yet they are not suitable for measuring the style similarity.

The basic idea of these metrics is to compare the predicted code snippet against the ground truth by calculating the intersection of contiguous sequences of code tokens (i.e., n-grams). 
It is recognized that due to the syntactic verbosity and coding conventions inherent to programming languages, two code snippets frequently share numerous n-grams that are incapable of reflecting their stylistic nuances.
Besides, the score is calculated based on the exact match of n-grams, which can be deemed excessively rigid. For example, compared with \texttt{token\_count}, \texttt{word\_count} is expected to be more stylistically similar to \texttt{words\_count}. However, both of them will be assigned a score of 0 under 2-gram match.

CSSim addresses style from three perspectives: variable naming, API invocation, and code structure. Variable naming is generally believed as a reflection of the programmer's preference. For API invocation, similar APIs often exist in various libraries or packages, the choice of APIs also indicates the preference. As for code structure, sometimes the swap of two lines does not influence the operation hence the order should also be considered. Besides, CSSim is calculated based on a softer measurement, edit distance \cite{editdistance}.

API invocation and variable name follow the same process. Here we take the variable name as an example.
We first extract all the variables from the code snippet to get $\mathbf{V} =\{v_i\}_{i=1}^N$. For each variable in the set, we find the most similar variable from the other code and take the edit distance between the two as the similarity of this variable. Then, we take the weighted average value of all the variables as the style distance in variable naming. The whole process can be described as:

\begin{equation}
    \mathrm{Dis_{V_1}}=\frac{1}{||\lambda||_1}\sum_{v_i \in \mathbf{V}_1} \lambda_{i} \min_{v_j \in \mathbf{V}_2} \mathrm{ED} (v_i, v_j),
\end{equation}
where $\mathbf{V}_1$ and $\mathbf{V}_2$ are extracted variables from two codes and $\mathrm{ED}$ denotes Edit Distance. $\lambda_{i}$ is normalized inverse document frequency (IDF) because we intend to decrease the impact of common words. 
To ensure symmetry in this metric, we update code distance in variable naming as:

\begin{equation}
    \mathrm{Dis_{Var}} = \frac{\mathrm{Dis_{V1}} + \mathrm{Dis_{V2}}}{2}.
\end{equation}

For the measurement of code structure, we simply apply Tree Edit Distance (TED) \cite{treeeditdistance} to the Abstract Syntax Tree transformed from the code. Similar to edit distance, TED quantifies the least amount of basic operations (Insertion, Deletion, and Substitution) required to transform one tree into the other. To calculate CSSim, we first calculate the Code Style Distance $\mathrm{CSDis}$ between two codes $c_1$ and $c_2$, which is:

\begin{equation}
    \mathrm{CSDis}(c_1, c_2) = \frac{\mathrm{Dis_{Var}}+\mathrm{Dis_{API}}+\mathrm{TED}}{3},
\end{equation}
where $\mathrm{Dis_{Var}}$, $\mathrm{Dis_{API}}$, $\mathrm{TED} \in [0, 1]$ hence $\mathrm{CSDis} \in [0, 1]$. We define $\mathrm{CSSim} = 1-\mathrm{CSDis}$.

\section{Experimental Setups}

\paragraph{Datasets} We evaluate ReCo across various search scenarios and programming languages: online forum StackOverflow CoNaLa \cite{conala}, coding challenge competence APPS \cite{apps}, general programming problems MBPP \cite{mbpp} and MBJP \cite{mbjp}. The first three datasets are written in Python while the last one is written in Java. The statistics of the datasets are shown in Appendix~\ref{app:dataset_stat}. We take the widely adopted Mean Reciprocal Rank (MRR) as the evaluation metric \cite{softinfonce}. MRR is the average of reciprocal ranks of the true code snippets for the given query.

\paragraph{Baselines} We apply ReCo on several models: \textbf{BM25}, an enhanced version of TF-IDF, is a statistical measure that matches certain keywords in codes with the given query. \textbf{CodeBERT} is a bi-modal model pre-trained on Masked Language Modeling and Replaced Token Detection. \textbf{UniXcoder} unifies both generating and understanding pre-training tasks to further enhance code representation learning by leveraging cross-modal contents. \textbf{Contriever} \cite{contriever} is an unsupervised dense information retrieval model that leverages contrastive learning for its training. \textbf{CodeT5+} is pre-trained on both unimodal code data and bimodal code-text data with a diverse set of pretraining tasks including span denoising, causal language modeling, contrastive learning, and text-code matching.

\begin{table}[t]
\centering
\resizebox{\columnwidth}{!}{%
\begin{tabular}{@{}lllll@{}}
\toprule
                      & CoNaLa        & MBPP          & APPS          & MBJP          \\ \midrule
\multicolumn{5}{l}{\textit{Unsupervised}} \\ 
BM25                  & 52.6          & 12.6          & 11.6          & 11.3          \\
\quad +\, GAR                  & 71.7          & 35.1          & 17.6          & 33.5          \\
\quad +\, ReCo                 & $\textbf{75.8}_{+4.1}$ & $\textbf{70.8}_{+35.7}$ & $\textbf{22.6}_{+5.0}$ & $\textbf{65.3}_{+31.8}$ \\ \midrule
UniXcoder             & 77.2          & 69.3          & 8.3           & 73.2          \\
\quad +\, GAR                  & 83.9          & 85.0          & 13.2          & 80.0          \\
\quad +\, ReCo                 & $\textbf{85.1}_{+1.2}$ & $\textbf{92.4}_{+7.4}$ & $\textbf{28.8}_{+15.6}$ & $\textbf{87.6}_{+7.6}$ \\ \midrule
Contriever            & 55.7          & 55.3          & 9.6           & 37.0          \\
\quad +\, GAR                  & 75.0          & 71.3          & 14.0          & 62.3          \\
\quad +\, ReCo                 & $\textbf{77.9}_{+2.9}$ & $\textbf{87.4}_{+16.1}$ & $\textbf{41.6}_{+27.6}$ & $\textbf{76.6}_{+14.3}$ \\ \midrule
CodeT5+            & 73.7          & 59.4          & 7.6           & 67.7          \\
\quad +\, GAR                  & 80.3          & 77.7          & 10.2         & 79.2          \\
\quad +\, ReCo                 & $\textbf{80.8}_{+0.5}$ & $\textbf{89.4}_{+11.7}$ & $\textbf{29.9}_{+19.7}$ & $\textbf{84.0}_{+4.8}$ \\ \midrule \midrule
\multicolumn{5}{l}{\textit{Supervised}} \\
CodeBERT              & 83.6          & 79.6          & 25.1          & 79.6          \\
\quad +\, GAR                  & \textbf{88.6} & 87.7          & 29.3          & 84.1          \\
\quad +\, ReCo                 & $85.0_{-3.6}$          & $\textbf{92.3}_{+4.6}$ & $\textbf{51.2}_{+21.9}$ & $\textbf{89.1}_{+5.0}$ \\ \midrule
UniXcoder          & 84.8          & 81.2          & 24.3          & 81.6          \\
\quad +\, GAR                  & 85.9          & 89.0         & 34.5          & 85.6          \\
\quad +\, ReCo                 & $\textbf{87.1}_{+1.2}$ & $\textbf{94.2}_{+5.2}$ & $\textbf{58.1}_{+23.6}$ & $\textbf{90.5}_{+4.9}$ \\ \bottomrule
\end{tabular}}
\caption{
Comparative analysis of various models w.r.t MRR(\%) when utilizing GAR or ReCo.}
\label{tab:overall}
\end{table}

\paragraph{Compared Metrics} We compare Code Style Similarity with several metrics used for measuring semantic similarity. \textbf{BLEU} measures how many words are shared between the generated and the reference sentence based on the modified n-gram precision. \textbf{ROUGE-L} computes the longest common subsequence of words. \textbf{CodeBLEU} is tailored for code snippets by setting higher weights for programming language keywords and considering data-flow and AST match as well.

\begin{figure*}[t]
\centering
\includegraphics[width=1.99\columnwidth]{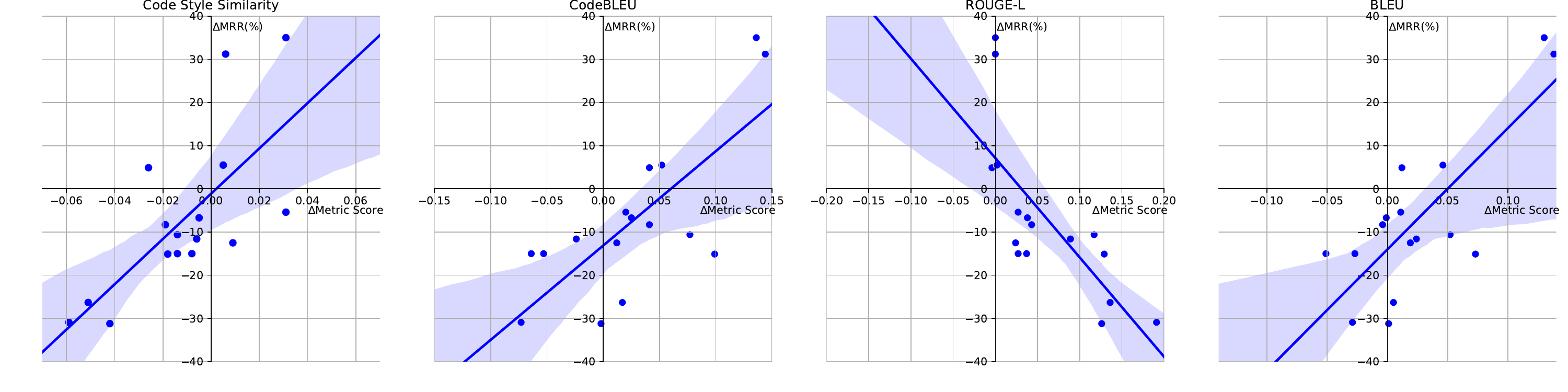}
\caption{Regression plots between $\Delta \mathrm{MRR}(=\mathrm{MRR}_{\mathrm{ReCo}}-\mathrm{MRR}_{\mathrm{GAR}})$ and $\Delta \mathrm{Metric Score}(=\mathrm{Metric}(c_q, c_c)-\mathrm{Metric}(c_q, c))$ under different evaluation metrics. The data points are from BM25 results on four datasets with four LLMs.} 
\label{fig:csd}
\end{figure*}

\paragraph{Implementation Details}
When prompting LLMs, we randomly sample 4 in-context examples from the training sets and set a temperature of 1. And when we prompt LLMs multiple times for the same input, each time we resample the in-context example. The maximum length of output for code summarization and generation is 128 and 256, respectively. For sparse retrieval, we use the default implementation from Pyserini \cite{pyserini}. For dense retrieval, during training, we adopt the default hyperparameters described in the original paper. They are trained for 10 epochs with a batch size of 32. Experiments are conducted on a Nvidia Tesla A100 GPU. Please refer to Appendix~\ref{sec:implementation} for more details.

\section{Results}

\paragraph{Overall Results} The results are shown in Table~\ref{tab:overall}. It is worth noting that our experiments encompass the use of various LLMs and multiple instances of both exemplar codes and rewritten codes for conducting ablation studies. Here we report the best performance when equipped with ReCo.  Comprehensive results can be found in Appendix~\ref{app:fullresult}. We can observe that ReCo significantly outperforms GAR on both supervised and unsupervised models across diverse search scenarios. With ReCo, the non-neural model BM25 can have competitive performance compared to neural models under zero-shot setting. And ReCo could boost the performance of zero-shot neural models similar to supervised models. We also evaluate ReCo on Contriever, a passage retrieval model that is not specifically trained for code-related tasks. We argue that ReCo can also benefit general-purpose retrieval models. Note that compared with GAR, ReCo does not bring any additional computation in real-time search because the rewritten code could be pre-processed and stored in the codebase.

\paragraph{Comparison among Evaluation Metrics}
To demonstrate the superiority of Code Style Similarity, we prove that \textit{existing metrics are not effective in measuring the style similarity between two codes} by contradiction. If existing metrics are effective, they should satisfy two necessary conditions: 1) the variation of metric scores $\Delta\mathrm{Metric Score}$ between $\mathrm{Metric}(c_q, c_c)$ and $\mathrm{Metric}(c_q, c)$ is in the same direction with code search performance gap between ReCo and GAR ($\Delta \mathrm{MRR}=\mathrm{MRR}_{\mathrm{ReCo}}-\mathrm{MRR}_{\mathrm{GAR}}$). This is because once the rewritten code is closer to the exemplar code in style, the code search performance should improve accordingly. 2) If we only choose the best one with the highest $\mathrm{Metric}(c_q, c)$ among multiple exemplar codes, the performance should be significantly better than randomly selecting one exemplar code. 

\begin{table}[t]
\centering
\resizebox{\columnwidth}{!}{%
\begin{tabular}{@{}lccccc@{}}
\toprule
\multirow{2}{*}{Dataset} & \multirow{2}{*}{Random} & \multicolumn{4}{c}{w/ the best exemplar code} \\ \cmidrule(l){3-6} 
                         &                         & CSSim    & CodeBLEU    & ROUGE-L    & BLEU    \\ \midrule
CoNaLa                   & 41.3                    & \textbf{43.6}     & 39.6        & 41.5       & 42.3    \\
MBPP                     & 24.0                    & \textbf{26.8}     & 25.1        & 23.7       & 24.0    \\
APPS                     & 14.1                    & \textbf{15.2}     & 14.8        & 14.2       & 14.2    \\
MBJP                     & 26.6                    & \textbf{29.9}     & 29.1        & 27.2       & 28.8    \\ \bottomrule
\end{tabular}%
}
\caption{Performance comparison of the best exemplar code selection versus random selection for GAR across four datasets, using the Code Llama-7B model.}
\label{tab:filter}
\end{table}

For the first condition, we analyze the results from BM25 on the four datasets with different LLMs including GPT3.5, Code Llama-7B, 13B, and 34B. Here we do not take the results from dense retrieval systems because neural models can capture the potential relationship among similar tokens. The numerical scores under different evaluation metrics are shown in Appendix~\ref{app:fullresult}. Regression plots are shown in Fig.~\ref{fig:csd}. According to our first condition, $\Delta\mathrm{Metric Score}$ and $\Delta \mathrm{MRR}$ should be consistent, which means that points are expected to be scattered on Quadrant I and III. We can see that most of the points in CodeBLEU, ROUGE-L, and BLEU are scattered on Quadrant IV. In other words, when the rewritten codes are considered to be more similar to the exemplar code by these metrics, the performance of ReCo, on the contrary, drops compared with GAR. Points from Code Style Similarity mostly fall on Quadrant I and III and the regression line nearly passes through the origin.

For the second condition, we analyze the results from BM25 equipped with GAR. For each query, we calculate the metric score between its exemplar codes and the true code in the codebase and then select the one with the highest metric score. The performance on four datasets after selecting the best exemplar code generated by Code Llama-7B is shown in Table~\ref{tab:filter}. We can observe that compared with random selection, the improvement brought by CodeBLEU, ROUGE-L, and BLEU is not significant generally. On the contrary, Code Style Similarity outperforms other settings. To better understand the preference of different evaluation metrics, we also conduct case studies in Appendix~\ref{app:casestudy}.

In conclusion, our findings indicate that existing metrics for measuring code style similarity fall short when subjected to two contradictory conditions. Conversely, Code Style Similarity (CSSim) demonstrably satisfies these criteria, highlighting its superior effectiveness in quantifying stylistic similarities in code. Furthermore, we observe a clear positive correlation between code style similarity as measured by CSSim and improvement in MRR for code search, thereby validating ReCo’s motivation that style normalization is advantageous.

\begin{table}[t]
\centering
\resizebox{\columnwidth}{!}{%
\begin{tabular}{@{}lcccc@{}}
\toprule
                        & CoNaLa & MBPP & APPS & MBJP \\ \midrule
UniXcoder             & 77.2          & 69.3          & 8.3           & 73.2          \\
UniXcoder\,+\,ReCo          & \textbf{85.1}   & \textbf{86.2} & \textbf{27.3} & \textbf{83.4} \\
\quad w/o original query\&code & 83.1   & 84.4 & 26.8 & 78.1 \\ \midrule
UniXcoder-ft          & 84.8          & 81.2          & 24.3          & 81.6          \\
UniXcoder-ft\,+\,ReCo       & \textbf{87.1}   & \textbf{88.0} & \textbf{48.8} & \textbf{85.4} \\
\quad w/o original query\&code & 85.5   & 84.8 & 39.0 & 80.3 \\ \bottomrule
\end{tabular}%
}
\caption{Comparative performance analysis of using exclusively LLM-generated codes versus a combination of LLM-generated codes, original queries, and codes. ``UniXcoder-ft'' represents UniXcoder after fine-tuning.
}
\label{tab:onlygen}
\end{table}

\paragraph{Using only LLM-generated codes}
To further demonstrate that the exemplar code and rewritten code are similar in style, we conduct experiments to only use these LLM-generated codes in the retrieval system. In other words, we use exemplar code to retrieve rewritten codes. The results of UniXcoder under fine-tuning and zero-shot settings on four datasets are shown in Table~\ref{tab:onlygen}. We can see from the results that only using LLM-generated codes can reach competitive performance compared with additionally using original queries and codes, and even outperform the setting only using original queries and codes, which indicates that the consistent style in exemplar code and rewritten code have made retrieval easier.

\paragraph{Impact of Different LLMs}

\begin{table}[t]
\centering
\resizebox{\columnwidth}{!}{%
\begin{tabular}{@{}lcccc@{}}
\toprule
                  & CoNaLa        & MBPP          & APPS          & MBJP          \\ \midrule
BM25              & 52.6          & 12.6          & 11.6          & 11.3          \\
\quad w/ Code Llama-7B  & 14.2          & 14.0          & 7.8           & 15.4          \\
\quad w/ Code Llama-13B & 29.8          & 26.3          & 8.1           & 28.4          \\
\quad w/ Code Llama-34B & 20.4          & 14.7          & 4.7           & 15.8          \\
\quad w/ GPT3.5         & \textbf{75.8} & \textbf{70.8} & \textbf{22.6} & \textbf{65.3} \\ \bottomrule
\end{tabular}%
}
\caption{Performance of ReCo on BM25 when using different LLMs to generate exemplar and rewritten codes.}
\label{tab:llm}
\end{table}

We explore the effect of using different LLMs in ReCo. The results on BM25 are shown in Table~\ref{tab:llm}. The full results including other retrieval models are in Appendix~\ref{app:fullresult}. GPT3.5's number of parameters is not released publicly but is estimated at around 175 billion. Generally, we observe that larger LLMs yield greater improvements. 
However, a decrement in performance is noted with the application of Code Llama-34B. This decrement is attributed to the model’s propensity to generate code not only for the prompted fifth example but also for the initial four in-context examples. Consequently, the generated code is often truncated due to output length limitations.

\paragraph{Impact of Number of Generated Codes}

\begin{figure}[t]
\centering
\includegraphics[width=0.99\columnwidth]{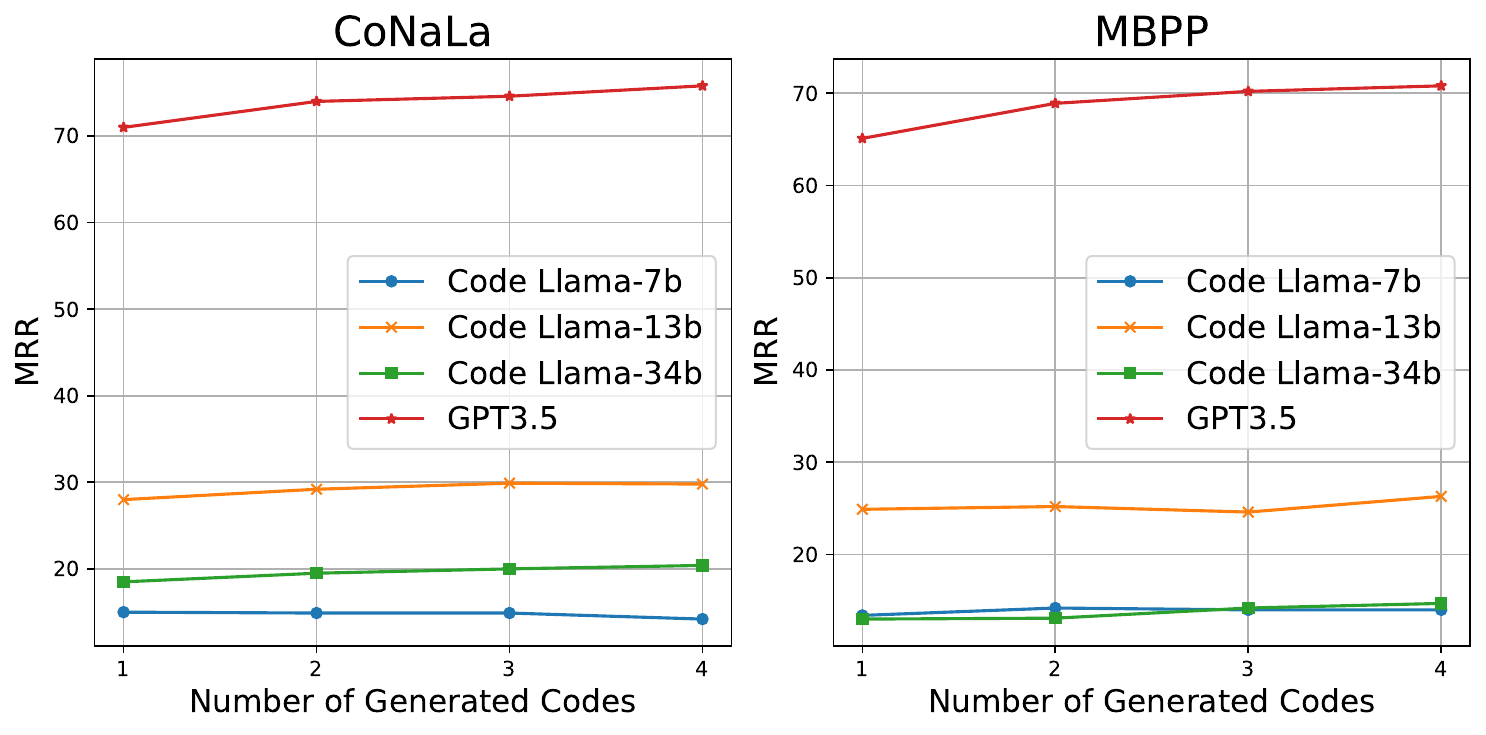}
\caption{Performance of BM25 + ReCo with different numbers of generated codes.} 
\label{fig:numgen}
\end{figure}

We also explore the effect of the numbers of generated exemplar codes and rewritten codes in ReCo. The outcomes of BM25 on CoNaLa and MBPP are depicted in Fig.~\ref{fig:numgen}, while a comprehensive compilation of results, inclusive of other retrieval models and datasets, can be found in Appendix~\ref{app:fullresult}. 
Our observations indicate a marginal enhancement when LLMs are tasked with generating more codes. We discern that the multiple codes generated exhibit similarities, with minor variations, attributable to the self-consistent style of each LLM. 
To address this, our future work will investigate controlled prompts that steer LLMs towards generating code with controlled stylistic variations, thereby enhancing the diversity of code generation.
Fig.~\ref{fig:numgen} also illustrates that the improvement tends to diminish as the quantity of generated codes increases.
In practical applications, it is essential to weigh the trade-off between performance enhancement and the incremental costs associated with generation.

\section{Broader Impact}
As we stated, the key motivation behind ReCo is to normalize the code style between exemplar code and original code. Indeed, there exist code normalization methods, but they only focus on superficial formatting such as the usage of indentation and naming convention (e.g., from camelCase to snake\_case). In this paper, we discuss code style normalization from a deeper perspective, implementation logic, and preference for variable naming or API invocation. We believe that this task has great potential as it could not only benefit code search but also many other code-related tasks like code review and code translation.

In this paper, we adopt LLMs to achieve the goal of style normalization by first summarizing the code snippet and then generating code based on the summary. This is because we find directly asking LLMs to rewrite the code results in very similar outputs. In the process of summarize-then-generate, models are expected to have great code intelligence hence there is no loss of information, as described in the theoretical insights of ReCo. Yet we are aware of the huge cost brought by LLMs. To decrease the cost, one promising solution is to train models specifically used for code style normalization. These models are considered to have much fewer parameters since much general knowledge in LLMs is not needed. To push forward the research of code style normalization, we propose a suitable evaluation metric, dubbed Code Style Similarity. In our future work, we plan to train such models to improve the efficiency of ReCo.

\section{Conclusion}
In this paper, we propose ReCo, an LLM-augmented code search framework built on GAR, that additionally rewrites the code in the code base to normalize the code style between exemplar code and code in the codebase. We evaluate ReCo on several code search models across various search scenarios with different programming languages. Experimental results demonstrate the effectiveness of ReCo by significantly boosting the performance of models. To encourage further research works on code style normalization and explain the effect of ReCo, we propose an evaluation metric Code Style Similarity. In our future work, based on this metric, we may develop new models that can more efficiently normalize the code style. 

\section*{Acknowledgement}
We thank the anonymous reviewers for their helpful comments and suggestions. 

\section*{Limitations}
There are mainly two limitations of this work. First, although ReCo does not require any additional computation in real-time search compared with GAR, both GAR and ReCo rely on the real-time generation of exemplar codes. Therefore, ReCo and GAR may have limitations when applied to tasks that demand low latency. The latency of generating exemplar codes depends on the time cost of LLM inference. As stated in research works focusing on GAR, over the years the cost of hardware has decreased a lot and there are many works proposed to improve the inference efficiency of LLMs \cite{hyde,query2doc}. We believe the efficiency problem of GAR and ReCo will be addressed in the future. The second limitation is that we do not evaluate ReCo on some extremely large-scale codebases like CodeSearchNet \cite{codesearchnet}. This is due to the time burden of generating exemplar codes and rewriting codes. For example, according to our estimation, there are 1,005,474 queries in total in CodeSearchNet hence generating one exemplar code for them costs more than two months. To address this limitation, we evaluate ReCo on several search scenarios covering coding challenge competence, online programming community, and general programming problems to show the effectiveness of ReCo.

\bibliography{custom}
\bibliographystyle{acl_natbib}

\appendix

\section{Complete Prompt}
\label{app:prompt}
Table~\ref{gen_prompt} and Table~\ref{sum_prompt} are two complete prompt examples for generating exemplar codes and summarizing original codes, respectively. Note that in the second step of rewriting original codes, we also adopt the prompt structure of generating exemplar codes but replace the description at last with the summary. 

\section{Experiment Settings}
\subsection{Dataset Statistics}
\label{app:dataset_stat}
The dataset statistics are shown in Table~\ref{tab:dataset_stat}. The numbers here are pairs of queries and their true code snippet. Code search models are asked to distinguish the correct code from the codes from other pairs. Note that in the original APPS dataset, there are 4,284 and 3,515 pairs in the training and test set, respectively. Due to the huge cost of prompting LLMs, we randomly sample a subset for our evaluation.

\begin{table}[h]
\centering
\begin{tabular}{@{}lcccc@{}}
\toprule
Dataset & CoNaLa & MBPP & APPS & MBJP \\ \midrule
Train   & 2,379   & 373  & 1,000 & 374  \\
Test    & 500    & 500  & 1,000 & 500  \\ \bottomrule
\end{tabular}
\caption{Statistics of the dataset used in our experiment.}
\label{tab:dataset_stat}
\end{table}

\begin{figure}[h]
\centering
\includegraphics[width=0.98\columnwidth, trim=0 240 0 0, clip]{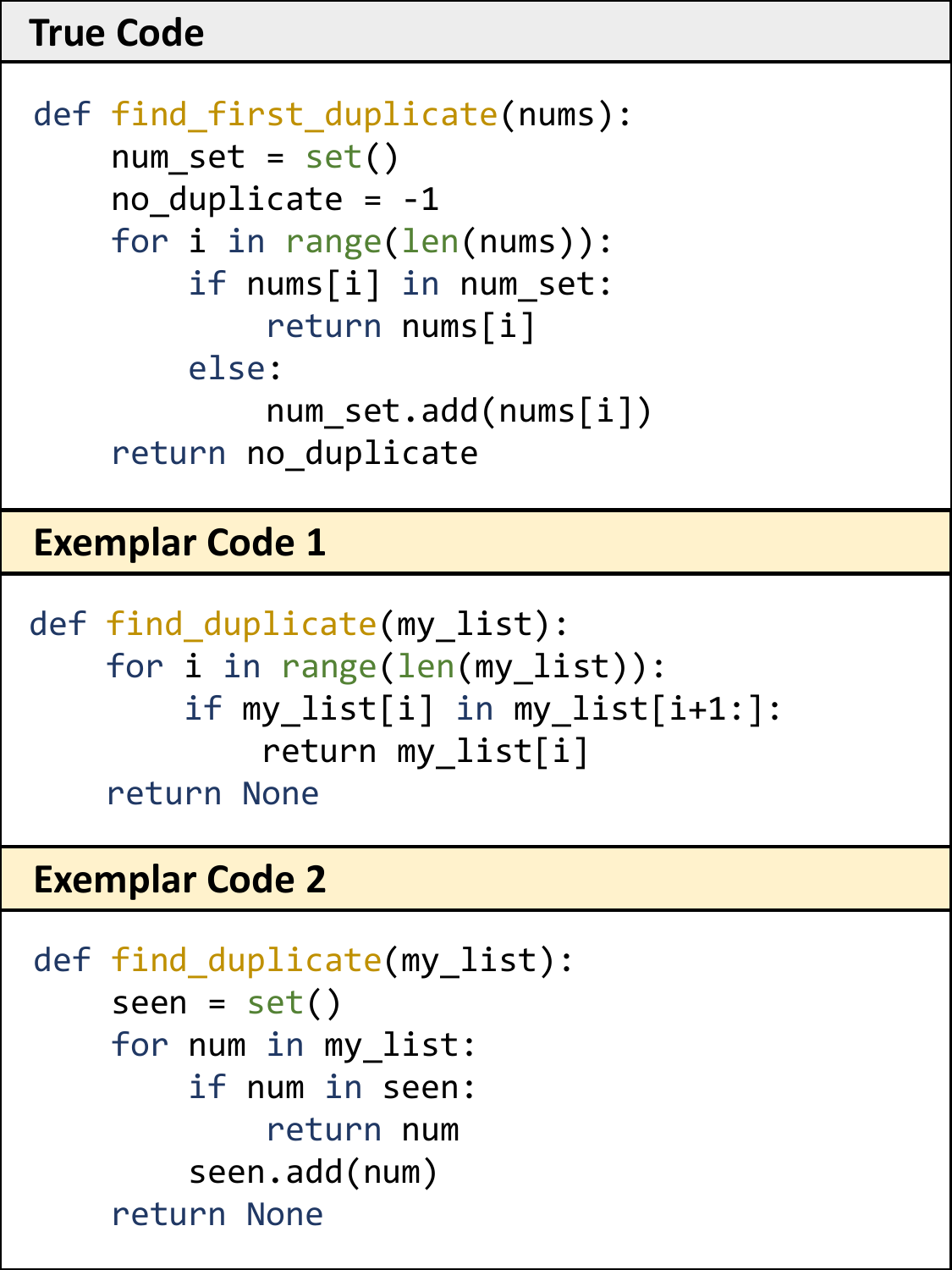}
\caption{Case study between Code Style Similarity and the existing metrics. The first exemplar code is preferred by existing metrics while the second one is preferred by Code Style Similarity. The second exemplar code is more similar to the true code from the perspective of style.}
\label{fig:casestudy}
\end{figure}

\subsection{MRR Calculation}
MRR is the average of reciprocal ranks of the true code snippets for the given query, which could be calculated as:

\begin{equation}
    \mathrm{MRR}=\frac{1}{|Q|}\sum_{i=1}^{|Q|}\frac{1}{\mathrm{Rank}_i}
\end{equation}
where $\mathrm{Rank}_i$ is the rank of the true code for the $i$-th given query $Q$.

\begin{table*}[]
\centering
\resizebox{\textwidth}{!}{%
\begin{tabular}{@{}llcccccccccccc@{}}
\toprule
\multirow{3}{*}{Metric} & \multirow{3}{*}{LLM} & \multicolumn{12}{c}{Datasets}                                                                                                                                                                                                                                                 \\ \cmidrule(l){3-14} 
                        &                      & \multicolumn{3}{c}{CoNaLa}                                        & \multicolumn{3}{c}{MBPP}                                          & \multicolumn{3}{c}{APPS}                                          & \multicolumn{3}{c}{MBJP}                                          \\ \cmidrule(l){3-14} 
                        &                      & M$(c_q,c_c)$ & M$(c_q,c)$ & $\Delta$M & M$(c_q,c_c)$ & M$(c_q,c)$ & $\Delta$M & M$(c_q,c_c)$ & M$(c_q,c)$ & $\Delta$M & M$(c_q,c_c)$ & M$(c_q,c)$ & $\Delta$M \\ \midrule
CSSim                   & CodeLlama-7B         & 0.492             & 0.543          & -0.051                       & 0.504             & 0.518          & -0.014                       & 0.493             & 0.498          & -0.005                       & 0.529             & 0.52           & 0.009                        \\
                        & CodeLlama-13B        & 0.534             & 0.593          & -0.059                       & 0.547             & 0.565          & -0.018                       & 0.481             & 0.5            & -0.019                       & 0.52              & 0.528          & -0.008                       \\
                        & CodeLlama-34B        & 0.522             & 0.564          & -0.042                       & 0.527             & 0.533          & -0.006                       & 0.526             & 0.495          & 0.031                        & 0.506             & 0.52           & -0.014                       \\
                        & GPT3.5               & 0.553             & 0.548          & 0.005                        & 0.553             & 0.522          & 0.031                        & 0.508             & 0.534          & -0.026                       & 0.502             & 0.496          & 0.006                        \\ \midrule
CodeBLEU                & CodeLlama-7B         & 0.16              & 0.143          & 0.017                        & 0.264             & 0.187          & 0.077                        & 0.153             & 0.128          & 0.025                        & 0.324             & 0.312          & 0.012                        \\
                        & CodeLlama-13B        & 0.116             & 0.189          & -0.073                       & 0.348             & 0.249          & 0.099                        & 0.171             & 0.13           & 0.041                        & 0.311             & 0.375          & -0.064                       \\
                        & CodeLlama-34B        & 0.181             & 0.183          & -0.002                       & 0.184             & 0.208          & -0.024                       & 0.151             & 0.131          & 0.020                        & 0.266             & 0.319          & -0.053                       \\
                        & GPT3.5               & 0.28              & 0.228          & 0.052                        & 0.385             & 0.249          & 0.136                        & 0.204             & 0.163          & 0.041                        & 0.462             & 0.318          & 0.144                        \\ \midrule
ROUGE-L                 & CodeLlama-7B         & 0.138             & 0.002          & 0.136                        & 0.118             & 0.001          & 0.117                        & 0.041             & 0.003          & 0.038                        & 0.025             & 0.001          & 0.024                        \\
                        & CodeLlama-13B        & 0.193             & 0.002          & 0.191                        & 0.13              & 0.001          & 0.129                        & 0.046             & 0.003          & 0.043                        & 0.038             & 0.001          & 0.037                        \\
                        & CodeLlama-34B        & 0.127             & 0.001          & 0.126                        & 0.0898            & 0.001          & 0.0888                       & 0.03              & 0.003          & 0.027                        & 0.027             & 0              & 0.027                        \\
                        & GPT3.5               & 0.007             & 0.005          & 0.002                        & 0.001             & 0.001          & 0                            & 0.003             & 0.007          & -0.004                       & 0                 & 0              & 0                            \\ \midrule
BLEU                    & CodeLlama-7B         & 0.017             & 0.012          & 0.005                        & 0.081             & 0.029          & 0.052                        & 0.017             & 0.018          & -0.001                       & 0.115             & 0.096          & 0.019                        \\
                        & CodeLlama-13B        & 0.026             & 0.055          & -0.029                       & 0.143             & 0.07           & 0.073                        & 0.019             & 0.023          & -0.004                       & 0.146             & 0.197          & -0.051                       \\
                        & CodeLlama-34B        & 0.026             & 0.025          & 0.001                        & 0.054             & 0.03           & 0.024                        & 0.021             & 0.01           & 0.011                        & 0.109             & 0.136          & -0.027                       \\
                        & GPT3.5               & 0.145             & 0.099          & 0.046                        & 0.195             & 0.065          & 0.13                         & 0.048             & 0.036          & 0.012                        & 0.309             & 0.171          & 0.138                        \\ \bottomrule
\end{tabular}%
}
\caption{Full results of $\mathrm{Metric Score}(c_q, c_c)$ and $\mathrm{Metric Score}(c_q, c)$ under different metrics. M is short for $\mathrm{MetricScore}$. $\Delta \mathrm{M}=\mathrm{M}(c_q, c_c)-\mathrm{M}(c_q, c)$.}
\label{tab:metricfull}
\end{table*}

\subsection{Implementation Details}
\label{sec:implementation}
For neural models, they all use the same set of hyperparameters. The maximum input length of codes and queries are both set to be 256. Models are trained by Adam and learning rate is set to 5e-6. We adopt mean pooling to get the representation of the whole input sentence to make sure the pooling mechanism is consistent with that during pre-training. The representations are normalized by the L2 norm. UniXcoder is initialized using the publicly available checkpoint at \url{https://huggingface.co/microsoft/unixcoder-base}, Contriever is initialized using \url{https://huggingface.co/facebook/contriever-msmarco}, and CodeT5+ is initialized using \url{https://huggingface.co/Salesforce/codet5p-110m-embedding}. CodeBERT is initialized using \url{https://huggingface.co/microsoft/codebert-base} and then pre-trained on the CodeSearchNet dataset \cite{codesearchnet} for 10 epochs. The pre-training setting is the same as in fine-tuning. All the experiments involving model training are running with 3 random seeds 1234, 12345, and 123456 and they all meet $p<0.01$ of significance tests.

\section{Case Study}
\label{app:casestudy}
We also conduct a case study to show the superiority of Code Style Similarity. Fig.~\ref{fig:casestudy} shows two exemplar codes generated by Code Llama-7B. The true code aims to find the first duplicate element in a given array of integers. Although both the two exemplar codes satisfy the description, their implementation style is different. The first exemplar code is preferred by CodeBLEU, ROUGE-L, and BLEU while the second one is preferred by Code Style Similarity. The true code uses a set to collect seen elements when traversing the list, which is also the logic in the second exemplar code. The first exemplar code implements the function in a different way by checking whether \texttt{my\_list[i]} appears in \texttt{my\_list[i+1:]}. We think the first code is preferred by existing metrics because lines 2-4 in the first exemplar code are very similar to lines 4-6 in the true code, which contributes a lot to the metric score.

\section{Full Results}
\label{app:fullresult}
In this section, we report the full experimental results. The results of $\mathrm{Metric Score}(c_q, c_c)$ and $\mathrm{Metric Score}(c_q, c)$ under different evaluation metrics are shown in Table~\ref{tab:metricfull}. The results of BM25 are shown in Table~\ref{tab:bm25full}. The results of fine-tuned UniXcoder and CodeBERT are shown in Table~\ref{tab:unixcoder-ft full} and Table~\ref{tab:codebert-ft full}, respectively. The results of UniXcoder, Contriever, and CodeT5+ under zero-shot setting are shown in Table~\ref{tab:neural zeroshot full}.

\begin{table*}[]
\centering
\begin{tabular}{@{}lccccccc@{}}
\toprule
\multirow{2}{*}{Model} & \multirow{2}{*}{LLM}           & \multirow{2}{*}{\#gen} & \multirow{2}{*}{Framework} & \multicolumn{4}{c}{Datasets}                                  \\ \cmidrule(l){5-8} 
                       &                                &                        &                            & CoNaLa        & MBPP          & APPS          & MBJP          \\ \midrule
\multirow{32}{*}{BM25} & \multirow{8}{*}{CodeLlama-7B}  & \multirow{2}{*}{1}     & GAR                        & 41.3          & 24.0          & 14.1          & 26.6          \\
                       &                                &                        & ReCo                       & 15.0          & 13.4          & 7.4           & 14.1          \\ \cmidrule(l){3-8} 
                       &                                & \multirow{2}{*}{2}     & GAR                        & 44.0          & 25.5          & 14.9          & 28.4          \\
                       &                                &                        & ReCo                       & 14.9          & 14.2          & 7.6           & 14.1          \\ \cmidrule(l){3-8} 
                       &                                & \multirow{2}{*}{3}     & GAR                        & 44.5          & 26.3          & 15.4          & 29.3          \\
                       &                                &                        & ReCo                       & 14.9          & 14.0          & 7.5           & 14.8          \\ \cmidrule(l){3-8} 
                       &                                & \multirow{2}{*}{4}     & GAR                        & 44.5          & 26.4          & 15.2          & 29.7          \\
                       &                                &                        & ReCo                       & 14.2          & 14.0          & 7.8           & 15.4          \\ \cmidrule(l){2-8} 
                       & \multirow{8}{*}{CodeLlama-13B} & \multirow{2}{*}{1}     & GAR                        & 58.9          & 40.0          & 16.0          & 41.8          \\
                       &                                &                        & ReCo                       & 28.0          & 24.9          & 7.7           & 14.1          \\ \cmidrule(l){3-8} 
                       &                                & \multirow{2}{*}{2}     & GAR                        & 62.7          & 42.1          & 16.5          & 4.4           \\
                       &                                &                        & ReCo                       & 29.2          & 25.2          & 8.0           & 28.1          \\ \cmidrule(l){3-8} 
                       &                                & \multirow{2}{*}{3}     & GAR                        & 64.0          & 41.4          & 16.8          & 45.1          \\
                       &                                &                        & ReCo                       & 29.9          & 24.6          & 8.0           & 28.6          \\ \cmidrule(l){3-8} 
                       &                                & \multirow{2}{*}{4}     & GAR                        & 63.9          & 42.2          & 17.0          & 45.4          \\
                       &                                &                        & ReCo                       & 29.8          & 26.3          & 8.1           & 28.4          \\ \cmidrule(l){2-8} 
                       & \multirow{8}{*}{CodeLlama-34B} & \multirow{2}{*}{1}     & GAR                        & 49.7          & 24.6          & 10.0          & 29.6          \\
                       &                                &                        & ReCo                       & 18.5          & 13.0          & 4.6           & 14.6          \\ \cmidrule(l){3-8} 
                       &                                & \multirow{2}{*}{2}     & GAR                        & 53.0          & 24.6          & 10.6          & 30.2          \\
                       &                                &                        & ReCo                       & 19.5          & 13.1          & 4.8           & 14.4          \\ \cmidrule(l){3-8} 
                       &                                & \multirow{2}{*}{3}     & GAR                        & 55.5          & 24.6          & 10.9          & 31.5          \\
                       &                                &                        & ReCo                       & 20.0          & 14.2          & 4.8           & 15.3          \\ \cmidrule(l){3-8} 
                       &                                & \multirow{2}{*}{4}     & GAR                        & 56.8          & 25.7          & 10.9          & 32.3          \\
                       &                                &                        & ReCo                       & 20.4          & 14.7          & 4.7           & 15.8          \\ \cmidrule(l){2-8} 
                       & \multirow{8}{*}{GPT3.5}        & \multirow{2}{*}{1}     & GAR                        & 65.5          & 30.2          & 16.3          & 30.6          \\
                       &                                &                        & ReCo                       & 71.0          & 65.1          & 21.2          & 61.8          \\ \cmidrule(l){3-8} 
                       &                                & \multirow{2}{*}{2}     & GAR                        & 69.7          & 34.5          & 17.0          & 32.2          \\
                       &                                &                        & ReCo                       & 74.0          & 68.9          & 21.9          & 65.4          \\ \cmidrule(l){3-8} 
                       &                                & \multirow{2}{*}{3}     & GAR                        & 71.0          & 35.3          & 17.3          & 33.1          \\
                       &                                &                        & ReCo                       & 74.6          & 70.2          & 22.8          & 65.6          \\ \cmidrule(l){3-8} 
                       &                                & \multirow{2}{*}{4}     & GAR                        & {\ul 71.7}    & {\ul 35.1}    & {\ul 17.6}    & {\ul 33.5}    \\
                       &                                &                        & ReCo                       & \textbf{75.8} & \textbf{70.8} & \textbf{22.6} & \textbf{65.3} \\ \bottomrule
\end{tabular}
\caption{Full results of BM25. \#gen denotes the number of generated and rewritten codes. \textbf{Bold} and \uline{underlined} results are the best performance of ReCo and the performance of GAR under the same setting, which are reported in Table~\ref{tab:overall}.}
\label{tab:bm25full}
\end{table*}

\begin{table*}[]
\centering
\begin{tabular}{@{}lccccccc@{}}
\toprule
\multirow{2}{*}{Model}      & \multirow{2}{*}{LLM}           & \multirow{2}{*}{\#gen} & \multirow{2}{*}{Framework} & \multicolumn{4}{c}{Datasets}                                  \\ \cmidrule(l){5-8} 
                            &                                &                        &                            & CoNaLa        & MBPP          & APPS          & MBJP          \\ \midrule
\multirow{20}{*}{UniXcoder} & \multirow{6}{*}{CodeLlama-7B}  & \multirow{2}{*}{1}     & GAR                        & 85.4          & 74.2          & 32.2          & 77.1          \\
                            &                                &                        & ReCo                       & 72.5          & 76.7          & 51.3          & 81.7          \\ \cmidrule(l){3-8} 
                            &                                & \multirow{2}{*}{2}     & GAR                        & 86.7          & 74.8          & 34.0          & 79.1          \\
                            &                                &                        & ReCo                       & 78.8          & 79.3          & 54.9          & 83.5          \\ \cmidrule(l){3-8} 
                            &                                & \multirow{2}{*}{3}     & GAR                        & 87.0          & 75.0          & {\ul 34.5}    & 79.3          \\
                            &                                &                        & ReCo                       & 81.7          & 80.2          & \textbf{58.1} & 84.1          \\ \cmidrule(l){2-8} 
                            & \multirow{6}{*}{CodeLlama-13B} & \multirow{2}{*}{1}     & GAR                        & 89.2          & 87.9          & 36.3          & 84.9          \\
                            &                                &                        & ReCo                       & 81.0          & 92.9          & 41.4          & 90.0          \\ \cmidrule(l){3-8} 
                            &                                & \multirow{2}{*}{2}     & GAR                        & 90.2          & 89.0          & 36.9          & 85.6          \\
                            &                                &                        & ReCo                       & 85.1          & 93.2          & 46.1          & 89.6          \\ \cmidrule(l){3-8} 
                            &                                & \multirow{2}{*}{3}     & GAR                        & 90.8          & {\ul 89.0}    & 38.3          & {\ul 85.6}    \\
                            &                                &                        & ReCo                       & 85.5          & \textbf{94.2} & 46.8          & \textbf{90.5} \\ \cmidrule(l){2-8} 
                            & \multirow{6}{*}{CodeLlama-34B} & \multirow{2}{*}{1}     & GAR                        & 87.0          & 81.3          & 29.9          & 81.2          \\
                            &                                &                        & ReCo                       & 82.3          & 65.2          & 28.4          & 66.9          \\ \cmidrule(l){3-8} 
                            &                                & \multirow{2}{*}{2}     & GAR                        & 87.6          & 83.8          & 31.2          & 83.6          \\
                            &                                &                        & ReCo                       & 85.6          & 75.7          & 34.5          & 72.8          \\ \cmidrule(l){3-8} 
                            &                                & \multirow{2}{*}{3}     & GAR                        & 88.4          & 83.6          & 32.9          & 84.6          \\
                            &                                &                        & ReCo                       & 87.0          & 78.8          & 38.2          & 76.6          \\ \cmidrule(l){2-8} 
                            & \multirow{2}{*}{GPT3.5}        & \multirow{2}{*}{1}     & GAR                        & {\ul 85.9}    & 79.9          & 44.5          & 80.5          \\
                            &                                &                        & ReCo                       & \textbf{87.1} & 88.0          & 48.8          & 85.4          \\ \bottomrule
\end{tabular}
\caption{Full results of UniXcoder after fine-tuning. \#gen denotes the number of generated and rewritten codes. \textbf{Bold} and \uline{underlined} results are the best performance of ReCo and the performance of GAR under the same setting, which are reported in Table~\ref{tab:overall}. Note that due to the cost of OpenAI's API for using GPT3.5, we only generate one exemplar code and rewrite the code once for the training set. And due to the GPU memory limit, we can set a maximum number of \#gen as 3.}
\label{tab:unixcoder-ft full}
\end{table*}

\begin{table*}[]
\centering
\begin{tabular}{@{}lccccccc@{}}
\toprule
\multirow{2}{*}{Model}     & \multirow{2}{*}{LLM}           & \multirow{2}{*}{\#gen} & \multirow{2}{*}{Framework} & \multicolumn{4}{c}{Datasets}                                  \\ \cmidrule(l){5-8} 
                           &                                &                        &                            & CoNaLa        & MBPP          & APPS          & MBJP          \\ \midrule
\multirow{20}{*}{CodeBERT} & \multirow{6}{*}{CodeLlama-7B}  & \multirow{2}{*}{1}     & GAR                        & 81.7          & 72.6          & 26.8          & 75.4          \\
                           &                                &                        & ReCo                       & 65.5          & 75.6          & 45.5          & 79.7          \\ \cmidrule(l){3-8} 
                           &                                & \multirow{2}{*}{2}     & GAR                        & 84.8          & 73.8          & 28.4          & 75.9          \\
                           &                                &                        & ReCo                       & 73.7          & 77.4          & 48.8          & 81.2          \\ \cmidrule(l){3-8} 
                           &                                & \multirow{2}{*}{3}     & GAR                        & 85.4          & 74.0          & {\ul 29.3}    & 76.1          \\
                           &                                &                        & ReCo                       & 77.1          & 78.6          & \textbf{51.2} & 81.2          \\ \cmidrule(l){2-8} 
                           & \multirow{6}{*}{CodeLlama-13B} & \multirow{2}{*}{1}     & GAR                        & 87.2          & 87.2          & 29.5          & 82.9          \\
                           &                                &                        & ReCo                       & 77.1          & 90.5          & 38.2          & 88.3          \\ \cmidrule(l){3-8} 
                           &                                & \multirow{2}{*}{2}     & GAR                        & 89.2          & 87.6          & 30.8          & 83.6          \\
                           &                                &                        & ReCo                       & 81.4          & 91.3          & 40.9          & 88.8          \\ \cmidrule(l){3-8} 
                           &                                & \multirow{2}{*}{3}     & GAR                        & 89.8          & {\ul 87.7}    & 31.6          & {\ul 84.1}    \\
                           &                                &                        & ReCo                       & 83.3          & \textbf{92.3} & 41.6          & \textbf{89.1} \\ \cmidrule(l){2-8} 
                           & \multirow{6}{*}{CodeLlama-34B} & \multirow{2}{*}{1}     & GAR                        & 86.1          & 78.5          & 23.3          & 79.7          \\
                           &                                &                        & ReCo                       & 78.3          & 59.6          & 22.2          & 61.0          \\ \cmidrule(l){3-8} 
                           &                                & \multirow{2}{*}{2}     & GAR                        & 87.4          & 81.1          & 25.2          & 81.8          \\
                           &                                &                        & ReCo                       & 82.4          & 68.9          & 27.0          & 68.4          \\ \cmidrule(l){3-8} 
                           &                                & \multirow{2}{*}{3}     & GAR                        & {\ul 88.6}    & 80.9          & 26.4          & 81.5          \\
                           &                                &                        & ReCo                       & \textbf{85.0} & 71.3          & 30.8          & 71.7          \\ \cmidrule(l){2-8} 
                           & \multirow{2}{*}{GPT3.5}        & \multirow{2}{*}{1}     & GAR                        & 83.3          & 81.6          & 38.2          & 80.9          \\
                           &                                &                        & ReCo                       & 82.4          & 83.5          & 43.1          & 81.4          \\ \bottomrule
\end{tabular}
\caption{Full results of CodeBERT after fine-tuning. \#gen denotes the number of generated and rewritten codes. \textbf{Bold} and \uline{underlined} results are the best performance of ReCo and the performance of GAR under the same setting, which are reported in Table~\ref{tab:overall}. Note that due to the cost of OpenAI's API for using GPT3.5, we only generate one exemplar code and rewrite the code once for the training set. And due to the GPU memory limit, we can set a maximum number of \#gen as 3.}
\label{tab:codebert-ft full}
\end{table*}

\begin{table*}[]
\centering
\begin{tabular}{@{}lccccccc@{}}
\toprule
\multirow{2}{*}{Model}      & \multirow{2}{*}{LLM}           & \multirow{2}{*}{\#gen} & \multirow{2}{*}{Framework} & \multicolumn{4}{c}{Datasets}                                  \\ \cmidrule(l){5-8} 
                            &                                &                        &                            & CoNaLa        & MBPP          & APPS          & MBJP          \\ \midrule
\multirow{8}{*}{UniXcoder}  & \multirow{2}{*}{CodeLlama-7B}  & \multirow{2}{*}{4}     & GAR                        & 77.2          & 62.5          & {\ul 13.2}    & 68.4          \\
                            &                                &                        & ReCo                       & 75.4          & 70.7          & \textbf{28.8} & 77.4          \\ \cmidrule(l){2-8} 
                            & \multirow{2}{*}{CodeLlama-13B} & \multirow{2}{*}{4}     & GAR                        & 85.1          & {\ul 85.0}    & 16.6          & {\ul 80.0}    \\
                            &                                &                        & ReCo                       & 82.5          & \textbf{92.4} & 25.4          & \textbf{87.6} \\ \cmidrule(l){2-8} 
                            & \multirow{2}{*}{CodeLlama-34B} & \multirow{2}{*}{4}     & GAR                        & 81.9          & 75.9          & 9.1           & 78.3          \\
                            &                                &                        & ReCo                       & 83.2          & 75.8          & 14.4          & 74.5          \\ \cmidrule(l){2-8} 
                            & \multirow{2}{*}{GPT3.5}        & \multirow{2}{*}{4}     & GAR                        & {\ul 83.9}    & 79.7          & 19.6          & 80.0          \\
                            &                                &                        & ReCo                       & \textbf{85.1} & 86.2          & 27.3          & 83.4          \\ \midrule
\multirow{8}{*}{Contriever} & \multirow{2}{*}{CodeLlama-7B}  & \multirow{2}{*}{4}     & GAR                        & 54.3          & 50.8          & {\ul 14.0}    & 44.8          \\
                            &                                &                        & ReCo                       & 55.0          & 66.2          & \textbf{41.6} & 67.2          \\ \cmidrule(l){2-8} 
                            & \multirow{2}{*}{CodeLlama-13B} & \multirow{2}{*}{4}     & GAR                        & 69.3          & {\ul 71.3}    & 16.4          & {\ul 62.3}    \\
                            &                                &                        & ReCo                       & 72.0          & \textbf{87.4} & 34.4          & \textbf{76.6} \\ \cmidrule(l){2-8} 
                            & \multirow{2}{*}{CodeLlama-34B} & \multirow{2}{*}{4}     & GAR                        & 61.3          & 60.3          & 10.1          & 50.1          \\
                            &                                &                        & ReCo                       & 68.3          & 63.9          & 20.3          & 56.0          \\ \cmidrule(l){2-8} 
                            & \multirow{2}{*}{GPT3.5}        & \multirow{2}{*}{4}     & GAR                        & {\ul 75.0}    & 65.6          & 17.9          & 58.7          \\
                            &                                &                        & ReCo                       & \textbf{77.9} & 79.5          & 24.1          & 72.4          \\ \midrule
\multirow{8}{*}{CodeT5+} & \multirow{2}{*}{CodeLlama-7B}  & \multirow{2}{*}{4}     & GAR                        & 67.4          & 61.3          & {\ul 10.2}    & 66.5         \\
                            &                                &                        & ReCo                       & 57.7          & 67.2          & \textbf{29.9} & 73.0         \\ \cmidrule(l){2-8} 
                            & \multirow{2}{*}{CodeLlama-13B} & \multirow{2}{*}{4}     & GAR                        & 78.6         & {\ul 77.7}    & 12.8         & {\ul 79.2}    \\
                            &                                &                        & ReCo                       & 72.8          & \textbf{89.4} & 24.7          & \textbf{84.0} \\ \cmidrule(l){2-8} 
                            & \multirow{2}{*}{CodeLlama-34B} & \multirow{2}{*}{4}     & GAR                        & 74.5          & 68.4          & 8.1          & 76.5          \\
                            &                                &                        & ReCo                       & 67.6          & 57.9          & 15.8          & 62.9          \\ \cmidrule(l){2-8} 
                            & \multirow{2}{*}{GPT3.5}        & \multirow{2}{*}{4}     & GAR                        & {\ul 80.3}    & 72.9          & 14.6         & 76.9         \\
                            &                                &                        & ReCo                       & \textbf{80.8} & 82.2          & 19.1          & 81.4          \\ \bottomrule
\end{tabular}
\caption{Full results of UniXcoder, Contriever, and CodeT5+ under zero-shot setting. \#gen denotes the number of generated and rewritten codes. \textbf{Bold} and \uline{underlined} results are the best performance of ReCo and the performance of GAR under the same setting, which are reported in Table~\ref{tab:overall}.}
\label{tab:neural zeroshot full}
\end{table*}

\begin{table*}[ht]
\centering
\scalebox{0.85}{\begin{tabular}{ll}
\toprule
Prompt        & \begin{tabular}[c]{@{}p{1.0\linewidth}@{}}\textbf{Please generate a python code snippet according to the last given description. Only output the code snippets. Do not explain the code. I will show you four examples first.}\\ \\ 
\textbf{Description:} Write a python function to find the index of an extra element present in one sorted array.\\ 
\textbf{Code:}\\
def find\_Extra(arr1,arr2,n) : \\
\hspace{15pt}    for i in range(0, n) : \\
\hspace{15pt}\hspace{15pt}        if (arr1[i] != arr2[i]) : \\
\hspace{15pt}\hspace{15pt}\hspace{15pt}            return i \\
\hspace{15pt}    return n
\\ \\ 
\textbf{Description:} Write a function to decode a run-length encoded given list.\\ 
\textbf{Code:} \\
def decode\_list(alist):\\
\hspace{15pt}    def aux(g):\\
\hspace{15pt}\hspace{15pt}        if isinstance(g, list):\\
\hspace{15pt}\hspace{15pt}\hspace{15pt}            return [(g[1], range(g[0]))]\\
\hspace{15pt}\hspace{15pt}        else:\\
\hspace{15pt}\hspace{15pt}\hspace{15pt}            return [(g, [0])]\\
\hspace{15pt}    return [x for g in alist for x, R in aux(g) for i in R]
\\ \\ 
\textbf{Description:} Write a python function to accept the strings which contains all vowels.\\ 
\textbf{Code:} \\
def check(string): \\
\hspace{15pt}  if len(set(string).intersection("AEIOUaeiou"))>=5: \\
\hspace{15pt}\hspace{15pt}    return ('accepted') \\
\hspace{15pt}else: \\
\hspace{15pt}\hspace{15pt}    return ("not accepted") 
\\ \\ 
\textbf{Description:} Write a function to compute binomial probability for the given number.\\ 
\textbf{Code:} \\
def get\_inv\_count(arr, n): \\
\hspace{15pt}	inv\_count = 0\\
\hspace{15pt}	for i in range(n): \\
\hspace{15pt}\hspace{15pt}		for j in range(i + 1, n): \\
\hspace{15pt}\hspace{15pt}\hspace{15pt}			if (arr[i] > arr[j]): \\
\hspace{15pt}\hspace{15pt}\hspace{15pt}\hspace{15pt}				inv\_count += 1\\
\hspace{15pt}	return inv\_count 
\\ \\
\textbf{Description:}Write a function to sort a given matrix in ascending order according to the sum of its rows. \\
\textbf{Code:} \end{tabular} \\ \midrule
Output & \begin{tabular}[c]{@{}p{1.0\linewidth}@{}}
def sort\_matrix\_by\_row\_sum(matrix):\\
\hspace{15pt}    row\_sums = []\\
\hspace{15pt}    for row in matrix:\\
\hspace{15pt}\hspace{15pt}        row\_sum = sum(row)\\
\hspace{15pt}\hspace{15pt}        row\_sums.append(row\_sum)\\
\hspace{15pt}    sorted\_matrix = [x for \_, x in sorted(zip(row\_sums, matrix))]\\
\hspace{15pt}    return sorted\_matrix
\end{tabular}  \\ \bottomrule
\end{tabular}}
\caption{A prompt example used for generating exemplar codes for the MBPP dataset. A more detailed prompt may increase the quality of the exemplar code and we leave this as our future work.}
\label{gen_prompt}
\end{table*}

\begin{table*}[ht]
\centering
\scalebox{0.85}{\begin{tabular}{ll}
\toprule
Prompt        & \begin{tabular}[c]{@{}p{1.0\linewidth}@{}}\textbf{What is the main purpose of the fifth python code snippet? Summarize in one sentence and be concise. I will show you four examples first.}\\ \\ 
\textbf{Code:}\\
def odd\_values\_string(str):\\
\hspace{15pt}  result = "" \\
\hspace{15pt}  for i in range(len(str)):\\
\hspace{15pt}\hspace{15pt}    if i \% 2 == 0:\\
\hspace{15pt}\hspace{15pt}\hspace{15pt}      result = result + str[i]\\
\hspace{15pt}  return result\\
\textbf{Purpose:} Write a python function to remove the characters which have odd index values of a given string.
\\ \\ 
\textbf{Code:} \\
from collections import defaultdict\\
def max\_aggregate(stdata):\\
\hspace{15pt}    temp = defaultdict(int)\\
\hspace{15pt}   for name, marks in stdata:\\
\hspace{15pt}\hspace{15pt}        temp[name] += marks\\
\hspace{15pt}    return max(temp.items(), key=lambda x: x[1])\\
\textbf{Purpose:} Write a function to calculate the maximum aggregate from the list of tuples.
\\ \\ 
\textbf{Code:} \\
def pos\_count(list):\\
\hspace{15pt}  pos\_count= 0\\
\hspace{15pt}  for num in list: \\
\hspace{15pt}\hspace{15pt}    if num >= 0: \\
\hspace{15pt}\hspace{15pt}\hspace{15pt}      pos\_count += 1\\
\hspace{15pt}  return pos\_count \\
\textbf{Purpose:} Write a python function to count positive numbers in a list.
\\ \\ 
\textbf{Code:} \\
import math\\
def volume\_tetrahedron(num):\\
\hspace{15pt}	volume = (num ** 3 / (6 * math.sqrt(2)))	\\
\hspace{15pt}	return round(volume, 2)\\
\textbf{Description:} Write a function to calculate volume of a tetrahedron.
\\ \\
\textbf{Code:} \\
def sort\_matrix(M):\\
\hspace{15pt}    result = sorted(M, key=sum)\\
\hspace{15pt}    return result\\
\textbf{Purpose:}
    \end{tabular} \\ \midrule
Output & \begin{tabular}[c]{@{}p{1.0\linewidth}@{}}
Write a function to sort a matrix (list of lists) based on the sum of each inner list.
\end{tabular}  \\ \bottomrule
\end{tabular}}
\caption{A prompt example used for summarizing the original codes for the MBPP dataset. A more detailed prompt may increase the quality of the rewritten code and we leave this as our future work.}
\label{sum_prompt}
\end{table*}

\end{document}